\documentclass[twocolumn,superscriptaddress,longbibliography,aps,pra,preprintnumbers]{revtex4-1}
\usepackage{graphicx}
\usepackage{bm}
\usepackage{color}
\usepackage{epstopdf}
\usepackage{amsmath}
\usepackage{amssymb}
\usepackage{epstopdf}
\usepackage[urlcolor=blue,colorlinks=true,citecolor=blue,
	linkcolor=blue,pdfstartview={FitH},bookmarks=false]{hyperref}
\usepackage{epsfig}
\usepackage[utf8]{inputenc}
\usepackage{dcolumn}
\usepackage{bm}
\usepackage{hyperref}

\begin{document}

\title{Leakage of Majorana mode into correlated quantum dot \\
       nearby its singlet-doublet crossover}

\author{T. Zienkiewicz}
\affiliation{Polish Air Force University, ul. Dywizjonu 303 no. 35, 08-521 Deblin, Poland}

\author{J. Bara\'nski}
\email[e-mail: ]{j.baranski@law.mil.pl}
\affiliation{Polish Air Force University, ul. Dywizjonu 303 no. 35, 08-521 Deblin, Poland}

\author{G. G\'{o}rski}
\affiliation{Faculty of Mathematics and Natural Sciences, University of Rzesz\'{o}w,
35-310 Rzesz\'{o}w, Poland}

\author{T. Doma\'{n}ski}
\email[e-mail: ]{doman@kft.umcs.lublin.pl}
\affiliation{Institute of Physics, M.\ Curie-Sk\l{}odowska University, 
20-031 Lublin, Poland}

\date{\today}

\begin{abstract}
We study quasiparticle spectrum of the correlated quantum dot deposited on superconducting 
substrate which is side-coupled to the Rashba nanochain, hosting Majorana end modes. 
Ground state of an isolated quantum dot proximitized to superconducting reservoirs 
is represented either by the singly occupied site or BCS-type superposition 
of the empty and doubly occupied configurations. Quantum phase transition between
these distinct ground states is spectroscopically manifested by the in-gap Andreev 
states which cross each other at the Fermi level. This qualitatively affects leakage 
of the Majorana mode from the side-attached nanowire. We inspect the spin-selective 
relationship between the trivial Andreev states and the leaking Majorana mode, 
considering (i) perfectly polarized case, when tunneling of one spin component 
is completely prohibited, and (ii) another one when both spins are hybridized 
with the nanowire but with different couplings.

\end{abstract}

\maketitle

\section{Introduction}
\label{sec:intro}

Recent development of the hybrid structures, comprising quantum 
dots (QDs) coupled the topologically superconducting nanowires 
\cite{Deng-2016,nichele.drachmann.17,Deng-2018,Wiesendanger-2018} 
provide new challenges going beyond mere observation of the Majorana 
bound states (MBS). Since energy levels of QDs in such hybrid systems 
are experimentally tunable one can inspect interplay of the topological 
states with the correlation effects and proximity-induced electron pairing.
Due to natural tendency of the Majorana modes to exist at boundaries of
finite size systems, one may expect their leakage into any side-attached 
quantum dot (QD) \cite{vernek.penteado.14} or more complex magnetic 
nanoislands \cite{DirkMorr_etal.18}.

It turns out, however, that efficiency of such process substantially 
depends on various parameters. For instance, in a weak coupling limit
the Majorana modes show up either by the constructive or
destructive interferometric lineshapes \cite{Gorski-2018}. True leakage 
of Majorana mode into nontopological region is possible only in strongly 
hybridized structures \cite{Hoffman-2017,Prada-2017,ptok.kobialka.17,
Szumniak-2017,Hoffman-2017} as indeed reported experimentally 
\cite{Deng-2016,nichele.drachmann.17,Deng-2018,Wiesendanger-2018}. 
But even under such conditions, leakage of the Majorana mode into 
the side-coupled QD region could be affected by various effects, e.g.\ 
by correlations. We address such issue here, studying an interplay 
between electron pairing and Coulomb repulsion  of the correlated QD 
proximitized to superconducting substrate where the quantum phase 
transition from the spinless (BCS-like) to the spinfull (singly 
occupied) configurations may occur. We analyze whether the Majorana 
mode would be able to leak into such configurations of the quantum dot.

The paper is organized as follows. We present the physical situation 
of our interest (Sec.~\ref{sec:setup}) and introduce the relevant 
microscopic model (Sec.~\ref{sec:model}).  
Next, we consider leakage of the Majorana mode into the  
quantum dot for the fully (Sec.~\ref{sec:polarized}) and
partly polarized (Sec.~\ref{sec:bothspins}) cases. Finally, we 
summarize the results (Sec.~\ref{sec:concl}) and give a brief reminder 
about the singlet-doubled transition in absence of Majorana mode
(Appendix~\ref{app:normal}).

\begin{figure}[t]
\includegraphics[width=1\columnwidth]{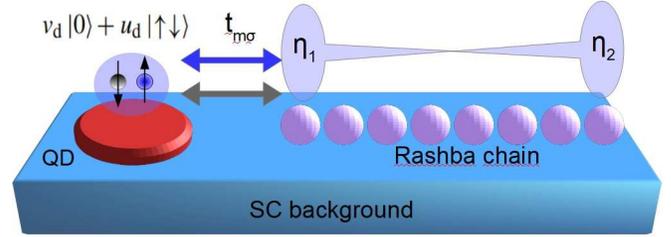}
\caption{Schematic view of the correlated quantum dot (QD) side-coupled to
the Rashba nanowire and both deposited on superconducting (SC) substrate,
where topological superconducting nanowire hosts two Majorana 
end-modes $\eta_{1}$ and $\eta_{2}$.} 
\label{scheme}
\end{figure}

\section{Physical setup}
\label{sec:setup}

We consider the strongly correlated quantum dot attached to the Rashba nanowire, 
both proximitized to the $s$-wave superconductor (SC). This setup 
(Fig.~\ref{scheme}) could be realized in the scanning tunneling microscopy 
(STM) measurements, using {\em Fe} or {\em Co} nanoscopic chains on the 
superconducting {\em Pb} \cite{Yazdani-2017}, {\em Al} or {\em Re} 
\cite{Wiesendanger-2018} substrates. 
%

In realistic situations the spin-orbit coupling along with the Zeeman 
effect break spin-rotational symmetry of the system. Such nanowire 
brought in contact with superconductor develops the intersite pairing 
of parallel spins. They are `tilted' with respect to $z$-axis, but one can 
project such pairing onto $\uparrow$ and $\downarrow$ components. For 
each of these sectors the intersite-pairing is characterized by 
effective amplitudes, mainly dependent on the applied magnetic field 
\cite{MaskaDomanski-2017}. 

Upon entering the topological phase, there appear two Majorana 
quasiparticles simultaneously in both spin channels but with 
different spectral weights, what has been indeed reported in 
the STM measurements using a ferromagnetic tip \cite{Yazdani-2017}. 
Such issue has been recently addressed by several groups 
\cite{Szumniak-2017,MaskaDomanski-2017,LiJian-2018}. 
J.~Klinovaja with co-workers \cite{Hoffman-2017} have shown that spin
up and down tunneling amplitudes between MBSs and quantum dot depend 
on the spin-orbit interaction length. Distance between the quantum dot 
and topological nanowire determines oscillations of the tunneling amplitudes.
By changing such dot-nanowire distance or varying the magnetic
field one can thus tune the hybrid system either to the fully or
to the partly polarized tunneling regimes. 

For perfect spin polarization of the QD-chain tunneling (e.g.\ $t_{\uparrow}\neq 0$,
$t_{\downarrow}=0$) one would expect signatures of the zero-energy mode 
to appear only in one spin channel (for $\uparrow$ electrons). However, 
the proximity induced on-dot pairing between opposite spins mixes both 
these channels. Some aspects of this situation have been addressed 
in Refs~\cite{Chirla-2016,Baranski-2017,Gorski-2018}. Since leakage of
the Majorana (zero-energy) mode is sensitive to electronic states near
the Fermi level, we would like to focus on crossing of the subg-gap (Andreev)
states caused by competition between on-dot pairing and Coulomb repulsion 
(which is spectroscopically manifested by the singlet-doublet quantum phase 
transition). We show that signatures of the Majorana mode look completely
different for both spin channels. In our considerations we take into 
account the case of (i) perfectly polarized tunneling of only one spin 
component while the other one is completely prohibited and (ii) another case 
where both spin electrons can be tunnel transferred, but with different amplitudes.

\section{Microscopic model}
\label{sec:model}

To inspect the mutual relationship between the Majorana mode and 
in-gap features of the correlated QD we restrict our attention 
to the limit of large energy gap $\Delta$ of 
the superconducting reservoir. Under such conditions  
the single level quantum dot is affected by the proximity-induced 
pairing. The proximitized QD is described by the Hamiltonian
$\sum_{\sigma} \epsilon d^{\dagger}_{\sigma} d_{\sigma}
+U n_{\downarrow}n_{\uparrow} + \frac{\Gamma_{S}}{2}(d_{\uparrow} 
d_{\downarrow}+d^{\dagger}_{\downarrow} d^{\dagger}_{\uparrow})$,  
where local pairing is represented by the pair source/sink terms.
We next consider coupling to the Majorana zero mode (MZM) 
represented by $\sum_{\sigma}\lambda \left( d^{\dagger}_{\sigma} 
\eta_{1} + \eta_{1} d_{\sigma}  \right)  + i\epsilon_m \eta_{1} \eta_{2}$. 
In our approach we recast the self-hermitian operators 
$\eta_{1,2}^{\dagger} = \eta_{1,2}$ by the standard fermionic ones  
\cite{Elliott-2015} $\eta_{1}=\frac{1}{\sqrt{2}}(f+f^{\dagger})$ and
$\eta_{2}=\frac{-i}{\sqrt{2}}(f-f^{\dagger})$. 
Summarizing, the effective low energy Hamiltonian can be expressed as 
\begin{eqnarray}
H^{\rm eff}_{QD} &\simeq & \sum_{\sigma} \epsilon d^{\dagger}_{\sigma} d_{\sigma}
+U n_{\downarrow}n_{\uparrow} + \frac{\Gamma_{S}}{2}(d_{\uparrow} 
d_{\downarrow}+d^{\dagger}_{\downarrow} d^{\dagger}_{\uparrow}) 
\label{sc_at_lim} \\
&+& \sum_{\sigma} t_{m \sigma} (d^{\dagger}_{\sigma} - d_{\sigma}) ( f + f^{\dagger} ) 
+ \epsilon_m \left( f^{\dagger} f 
+ \frac{1}{2} \right).
\nonumber
\end{eqnarray}
We assume that both spin components ($\sigma = \uparrow , \downarrow $) can be 
transferred between dot and Majorana host and $t_{m \sigma}=\lambda_{\sigma} / 
\sqrt{2}$ represent the hopping integrals of such processes. 

Our objective is to determine the Green's functions 
${\cal{G}}(\omega)=\langle \langle \Psi ; \Psi^{\dagger} \rangle \rangle$ 
defined in the Nambu matrix notation $\Psi^{\dagger}=(d_\uparrow,d_\uparrow^{\dagger},
d_\downarrow,d_\downarrow^{\dagger},f,f^{\dagger})$. The equation of 
motion technique applied to noncorrelated problem yields 
\begin{widetext}
\begin{eqnarray} 
\lim_{U=0} {\cal{G}}^{-1}(\omega) =
\left( \begin{array}{cccccc}  
\omega-\epsilon+i\Gamma_N/2 &0&0& \Gamma_S/2 & -t_{m\uparrow} & -t_{m\uparrow}\\
0&\omega+\epsilon+i\Gamma_N/2&-\Gamma_S/2& 0 & t_{m\uparrow} & t_{m\uparrow}\\
0&-\Gamma_S/2&\omega-\epsilon+i\Gamma_N/2 &0 & -t_{m\downarrow} & -t_{m\downarrow}\\
\Gamma_S/2 &0&0&\omega+\epsilon+i\Gamma_N/2 & t_{m\downarrow} & t_{m\downarrow} \\
-t_{m\uparrow} & t_{m\uparrow} &-t_{m\downarrow} & t_{m\downarrow}& \omega-\epsilon_{m}& 0\\
-t_{m\uparrow} & t_{m\uparrow} &-t_{m\downarrow} & t_{m\downarrow}& 0    & \omega+\epsilon_{m} 
\end{array}\right).
\label{Gr66}
\end{eqnarray} 
To account for the correlation effects one can numerically diagonalize 
8$\times$8 Hamiltonian matrix, determining the eigenenergies and 
transition elements between them. Another equivalent route can rely 
on the superconducting atomic limit solution \cite{Vecino-2003}, extending it
to the present model
\begin{eqnarray} 
{\cal{G}}^{-1}(\omega) =
\left( \begin{array}{cccccc}  
G(\omega) &0&0& F(\omega) & 0 &  0 \\
0&-G^{*}(-\omega)&F^{*}(-\omega)& 0 & 0 & 0\\
0&F^{*}(-\omega)& G(\omega) &0 & 0 & 0\\
F(\omega) &0&0&-G^{*}(-\omega) & 0 & 0 \\
0 & 0 & 0 & 0 & \frac{1}{\omega-\epsilon_{m}}& 0\\
0 & 0 & 0 & 0 & 0    & \frac{1}{\omega+\epsilon_{m}} 
\end{array}\right)^{-1} 
-
\left( \begin{array}{cccccc}  
0 & 0 & 0 & 0 & t_{m\uparrow} &  t_{m\uparrow}\\
0 & 0 & 0 & 0 & -t_{m\uparrow} & - t_{m\uparrow}\\
0 & 0 & 0 & 0 & t_{m\downarrow} & t_{m\downarrow}\\
0 & 0 & 0 & 0 & - t_{m\downarrow} & -t_{m\downarrow} \\
t_{m\uparrow} & -t_{m\uparrow} &t_{m\downarrow} & -t_{m\downarrow}& 0 & 0\\
t_{m\uparrow} & -t_{m\uparrow} &t_{m\downarrow} & -t_{m\downarrow}& 0 & 0 
\end{array}\right).
\label{full_solution}
\end{eqnarray} 
where
\begin{eqnarray}
{ G}(\omega) &=&   \frac{\alpha_s \;u_{d}^{2}}
{\omega-\left( \frac{U_{d}}{2}+E_{d} \right)} + \frac{\beta_s \; 
v_{d}^{2}}{\omega-\left( \frac{U_{d}}{2}-E_{d} \right)} 
+ \frac{\alpha_s 
\; v_{d}^{2}}{\omega+\left( \frac{U_{d}}{2}+E_{d} \right)} 
+ \frac{\beta_s \; u_{d}^{2}}{\omega+\left( \frac{U_{d}}{2}-E_{d} \right)},
 \label{G11_atomic} 
\\
{F}(\omega) &=&   \frac{\alpha_s\;u_{d}v_{d}}
{\omega-\left( \frac{U_{d}}{2}+E_{d} \right)}  - \frac{\beta_s \; 
u_{d}v_{d}}{\omega-\left( \frac{U_{d}}{2}-E_{d} \right)} 
 - 
 \frac{\alpha_s 
\; u_{d}v_{d}}{\omega+\left( \frac{U_{d}}{2}+E_{d} \right)} 
+ \frac{\beta_s \; u_{d}v_{d}}{\omega+\left( \frac{U_{d}}{2}-E_{d} \right)},
 \label{G12_atomic}
\end{eqnarray}
\end{widetext}
with the energy $E_{d} = \sqrt{(\epsilon+U/2)^2 + (\Gamma_{S}/2)^2}$, 
the usual BCS-type coefficients $u_{d}^{2} = \frac{1}{2} \left[ 1 + \frac{\epsilon+U/2}
{E_{d}} \right] =1 -v_{d}^{2}$
and the relative spectral weights 
\begin{eqnarray}
\alpha_s= \frac{e^{\beta U/2}+e^{-\beta E_d}}{2e^{\beta U/2}+e^{-\beta E_d} +e^{\beta E_d}}=1-\beta_s. 
\end{eqnarray} 
In what follows we will quasiparticle spectrum of the correlated quantum dot,
assuming that the topological nanowire is long enough so that any overlap
between the end-modes is negligible ($\epsilon_{m}\simeq 0$). 
 
\section{Fully polarized case}
\label{sec:polarized}

Let us  first consider the fully spin-polarized case when tunneling of one spin 
component, say $\downarrow$, is totally prohibited.  This situation could 
occur for very strong magnetic fields applied along the topological nanowire.
We show that combined effect of MZM leakage and proximity induced pairing 
give rise to zero modes apparent in energy spectrum of both spin 
components even if tunneling rate of one spin is turned off. 
We inspect the influence of these zero states on the characteristic 
features of a phase transition from the BCS-like singlet state ($S=0$) 
to the correlation-dominated doublet configuration. We show that even 
though zero modes appear in both spin channels, each of them have 
completely different character.

When a quantum dot or other nanoobject is tunnel coupled to the Rashba 
chain hosting Majorana particles some part of these zero-energy modes can 
be transferred to QD region. In consequence, the spectral function of QD 
reveals additional feature pinned to the Fermi level. For the perfectly 
polarized tunneling amplitude  ($t_{m \downarrow}=0$ and $t_{m \uparrow}\neq 0$) 
one would expect such effect to modify only the spectral function of the directly 
coupled spin component \cite{Lee-2013,Lopez-2014,Weymann-2017,Weymann-2017A}.  
However, due to the superconducting proximity effect the electrons (say $\uparrow$) 
leaking from nanochain into QD region are bound into local pairs with their 
opposite spin ($\downarrow$) partners \cite{Baranski-2017,Gorski-2018}. 
Thereby, the Majorana mode becomes apparent in the energy spectrum of 
electrons for which the direct tunneling to nanochain is prohibited. 
This feature in $\downarrow$ spin sector,  however, should be considered 
merely as a superconducting response for the leaking Majorana mode of 
the directly coupled spin $\uparrow$. Different nature of these zero-energy 
states is evident, both for the uncorrelated case  (Fig.\ \ref{fig1})
and in a behavior of the Andreev states of the correlated quantum 
dot near a quantum phase transition from the BCS-type singlet 
to the correlation driven doublet state (Fig. \ref{fig2}).  

\subsection{Energy spectrum of uncorrelated QD}
\label{sec:uncorrelated}

\begin{figure}[t]
\epsfxsize=9cm\centerline{\epsffile{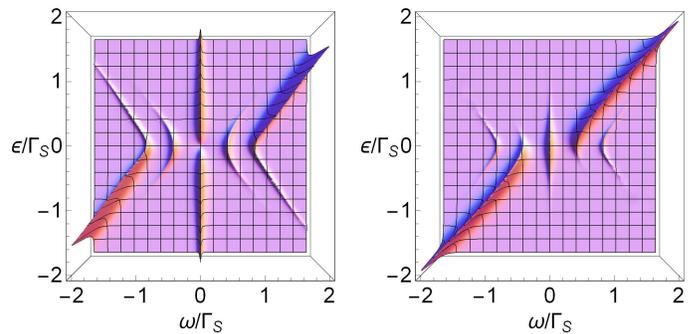}}
\caption{Density of states for spin $\uparrow$ (left) and $\downarrow$ (right) 
electrons obtained for $t_{m \uparrow}=0.6\Gamma_{S}$, $t_{m \downarrow}=0$ 
in absence of correlations ($U=0$).} 
\label{fig1}
\end{figure}

Let us focus first on the noncorrelated case, $U=0$.
By inspecting the matrix Green's function (\ref{Gr66}) in the limit 
$\Gamma_{N}\rightarrow 0$ we notice that it is characterized by five 
poles: zero-energy state, two Andreev bound states at $\pm \sqrt{\epsilon^2
+(\Gamma_S/2)^2}$ and another two `molecular' states $\pm \sqrt{\epsilon^2
+(\Gamma_S/2)^2+(2t_{m \uparrow})^2}$, resulting from hybridization of 
the Andreev bound states with the Rashba chain  \cite{Gorski-2018}. 
In Fig.\ \ref{fig1} we visualize qualitative differences between 
the spectral weights of $\uparrow$ and $\downarrow$ spin sectors  obtained
in the strong coupling limit $t_{\uparrow}=0.6 \Gamma_S$.

In the spectrum of $\uparrow$ electrons we observe the well pronounced 
zero-energy mode directly leaking from the proximitized Rashba chain in much the same
fashion as initially predicted for the quantum dot attached to the Kitaev wire
\cite{vernek.penteado.14}. Contrary to that, the opposite spin electrons are not 
directly coupled to the chain therefore zero-energy mode emerges as a consequence 
of the local pairing. Some tiny Majorana mode shows up in this $\downarrow$ spin
sector only when the QD energy $\epsilon$ is close to the Fermi level, otherwise 
it quickly fades away.  

Furthermore, one can notice a small dip in the zero-energy signature of spin 
$\uparrow$ electrons appearing near $\epsilon=0$. This comes from destructive 
feedback effect when the original QD level coincides with Majorana mode. Such 
destructive interference pattern has been described by our group 
\cite{Baranski-2017,Gorski-2018}. We have pointed out that even when spectral 
weight of the zero-energy mode in spectrum of directly coupled spins is suppressed, 
the zero-energy mode in opposite spin channel would be enhanced. 
As the QD-chain tunneling amplitude of spin $\downarrow$ electrons 
is turned off, spin $\downarrow$ electrons do not take part in destructive interference. 
Consequently in the left panel of Fig.\ \ref{fig2} we observe well shaped zero mode.  
It is worth noticing that spectral weights of original Andreev states 
($\pm \sqrt{\epsilon^2+(\Gamma_S/2)^2}$) in spectrum of $\uparrow$ electrons 
are reduced in comparison with the molecular ones. 
Contrary to that, for electrons with prohibited dot-chain tunneling we
observe that pure Andreev states are dominant.

\subsection{Majorana near singlet-doublet crossover}
\label{sec:QPT}

\begin{center}
\begin{figure*}
\epsfxsize=15cm\centerline{\epsffile{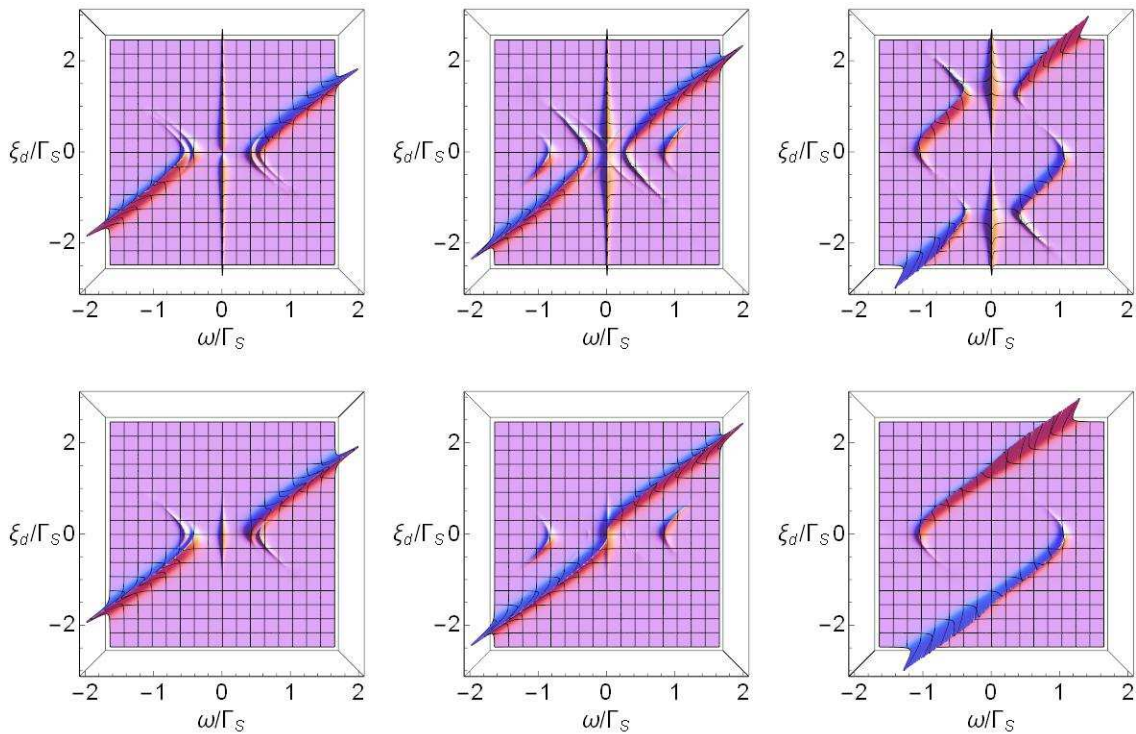}}
\caption{Density of states for spin $\uparrow$ (upper panel) and $\downarrow$ 
(bottom panel) electrons obtained for $t_{m \uparrow}=0.3\Gamma_{S}$, $t_{m \uparrow}=0$ 
and $U/\Gamma_S= 0$ (left panel), $U/\Gamma_S= 1$ (middle panel), $U/\Gamma_S= 3.5$ (right panel). 
Following Ref.\ \cite{Bauer_2007} we denote by $\xi_{d}=\epsilon-U/2$.}
\label{fig2}
\end{figure*}
\end{center}

We shall now discuss the correlation effects, driven by the Coulomb repulsion $U$.
In absence of the Rashba nanowire upon varying the ratio $U/\Gamma_{S}$ there appear 
the regular Andreev bound states at $\pm [U/2- \sqrt{(\epsilon+U/2)^2+(\Gamma_S/2)^2]}$ 
which eventually cross each other at the singlet-doublet (see Appendix 
\ref{app:normal} for an explanation of this quantum phase transition). 
Proximity induced zero-energy feature in the spin $\downarrow$ sector does 
not affect significantly this characteristic crossing, because it does not 
originate directly from the leaking Majorana mode. The strong Coulomb repulsion 
disfavors on-dot pairing, therefore upon increasing $U/\Gamma_{S}$ also the side 
effects of the local pairing are gradually suppressed. One of them is the zero-energy 
mode appearing in the spectrum of spin $\downarrow$ electrons, originating solely 
from electron pairing. For this reason, upon traversing the singlet-doublet 
quantum phase transition the zero-energy mode of spin $\downarrow$ electrons 
is gradually washed out from the strongly correlated regime (see right bottom 
lower panel in Fig.\ \ref{fig2}).   

In the case of spin $\uparrow$ electrons the spectrum looks completely different. 
Upon increasing the correlations strength the ABS states are split and form 
the molecular branches. These new states no longer cross each other. Approaching 
the singlet-doublet transition  we observe emergence of the zero-energy state 
(directly leaking from the Rashba nanochain) while the molecular Andreev states 
{\em repel} one another instead of the regular crossing (shown in Fig.\ 
\ref{figSAL}). Such avoided-crossing behavior (see the upper right panel 
in Fig.\ \ref{fig2}) can be regarded as additional signature, of the proximity 
induced Majorana mode that repels the `trivial' (finite-energy) Andreev states. 
This result generalizes the previously discussed topological/nontopological 
different nature of the bound states in hybrid structures comprising 
the uncorrelated  quantum dots attached to the topological superconducting 
wires hosting the Majorana modes
\cite{Prada-2017,Ptok-2017,Szumniak-2017,Hoffman-2017,Deng-2018}.
We hope that empirical detection would be feasible using the spin-polarized
STM measurements and varying the level $\epsilon$ of correlated QD by
the gate potential.

\begin{figure}
\epsfxsize=9cm\centerline{\epsffile{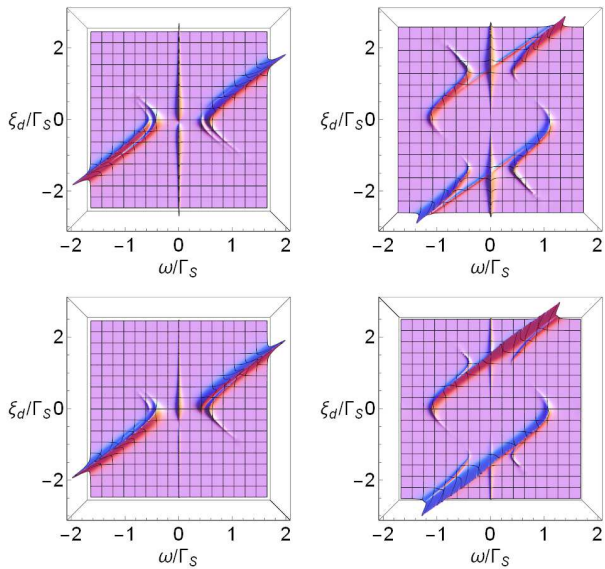}}
\caption{Density of states for spin up (upper panel) and spin down (bottom panel) electrons 
obtained for $t_{m \uparrow}=0.3\Gamma_{S}, t_{m \downarrow}=0.1\Gamma_{S}$ and $U/\Gamma_S= 0$ 
(left panel),  $U/\Gamma_S= 3.5$ (right panel).} 
\label{figspin}
\end{figure}

\section{Partial spin-polarization}
\label{sec:bothspins}

Theoretical studies indicate that for specific conditions, it would be possible to achieve 
nearly perfectly spin polarized tunneling between the quantum dot and chain \cite{Hoffman-2017}. 
However in realistic experimental setups both spin electrons can be transferred, although 
with substantially different tunneling amplitudes. In this subsection we briefly address 
such partially polarized case. 
Motivated by the spin polarized STM measurements \cite{Yazdani-2017} we assume 
in our approach the tunneling rate of spin $\downarrow$ electrons to be 
$t_{m \downarrow} = 0.1 \Gamma_{S}$ which is three times weaker than the coupling 
amplitude for spin $\uparrow$ electrons ($t_{m  \uparrow} =0.3 \Gamma_{S}$). 
Fig.\ \ref{figspin} compares the density of states of each spin electrons 
in the non-correlated (left panels) and strongly correlated regime (right panels), respectively. 

In the strongly correlated limit we can practically observe a convolution of 
the features typical for each spin sectors of the fully polarized case. Yet 
one can clearly resolve the {\em avoided-crossing} behavior in the dominant
$\uparrow$ spin coupling channel with the well pronounced Majorana mode
separating them. Spectrum of the opposite spin sector, on the other hand,
is predominantly reminiscent of the continuous Andreev band state branches
with only very residual signatures of the zero-energy modes. Correlations 
could thus be very useful for distinguishing the qualitatively different 
character of the spin-polarized spectra of the quantum dot.

\section{Summary}
\label{sec:concl}

We have studied the energy and spin-dependent spectra of the proximitized 
correlated quantum dot attached to the topologically superconducting nanochain, 
hosting the Majorana end-modes. We have analyzed influence of the correlations 
(responsible for a quantum phase transition from the spinless BCS-type to 
the spinful configuration) on efficiency of the Majorana mode leakage.
Our study predicts that the zero-energy states might appear in both spin 
sectors (due to the superconducting proximity effect), however their spectroscopic 
signatures are going to be qualitatively different. The spin channel which 
is directly coupled to the Majorana mode is characterized by (a) the well 
separated Andreev branches (of an avoided-crossing behavior) coexisting 
with (b) the zero-energy feature of a sizable spectral weight. The other 
spin sector, which is not directly coupled to the Majorana mode, 
in the strongly correlated limit is characterized by the continuous 
Andreev branches traversing the zero energy without any trace of 
the leaking Majorana mode. Such spin sector can eventually allow the Majorana
mode to appear but only in the weak correlation regime, when the proximitized
quantum dot stays in the BCS-type configuration.

Correlations can hence be very beneficial for spin-selective leakage
of the Majorona modes on the quantum dots. Such effects would be
easily detectable, using the polarized STM measurement analogous 
to those already reported in Ref. \cite{Yazdani-2017}. 

\begin{figure}[b!]
\epsfxsize=5.5cm\centerline{\epsffile{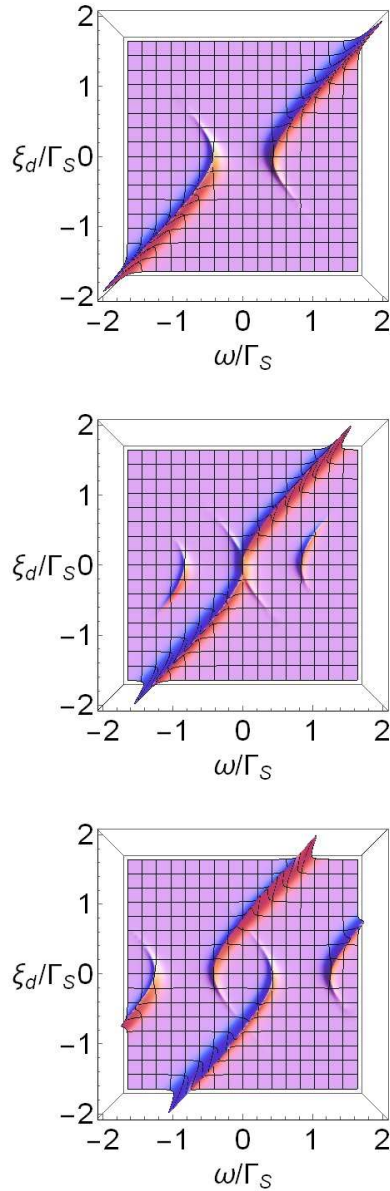}}
\caption{Spectral function $\rho(\omega)$ of the correlated quantum dot as a function 
of the energy $\xi_d=\epsilon+U/2$ obtained at low temperature $T=0.1\Gamma_{S}$ in 
the ‘superconducting atomic limit’ $\Gamma_{N} \rightarrow 0^{+}$ for several values 
of the Coulomb potential $U=0$ (upper panel) $U=\Gamma_{S}$ (middle panel), $U=2\Gamma_{S}$ 
(bottom panel).} 
\label{figSAL}
\end{figure}

\begin{acknowledgments}
This research has been conducted in the framework of the project ``Analysis 
of nanoscopic systems coupled with superconductors in the context of quantum 
information processing"  No.\ GB/5/2018/209/2018/DA  funded in period 
2018-2021 by the Ministry of National Defense Republic of Poland (J.B.,T.Z.).
This project has been also supported by National Science Centre (NCN, Poland) 
under the grants UMO-2018/29/B/ST3/00937 (G.G.) and UMO-2017/27/B/ST3/01911 (T.D.).
\end{acknowledgments}

\appendix
\section{Quantum phase transition in absence of Majorana mode}
\label{app:normal}

For illustration of the singlet-doublet phase transition we recall briefly 
the results obtained for the correlated quantum dot in absence of the Rashba chain. 
For this purpose we analyze the subgap Andreev states in the superconducting atomic 
following several earlier works \cite{Rodero-2012,Baranski-2013,Bauer_2007}. 
The true eigenstates are
represented by the doublet configurations $\left| \uparrow\right>$,
$\left| \downarrow \right>$ (corresponding to spin $S=\frac{1}{2}$) 
and the BCS-type singlet states $u_{d} \left| 0 \right> 
+ v_{d} \left| \downarrow \uparrow \right>$ or $-v_{d} \left| 0 \right>+ u_{d} 
\left| \downarrow \uparrow \right>$ ($S=0$). 
It can be shown that when correlations are weak $U<\Gamma_{S}$, 
the ground state is represented by singlet state in full range of 
initial dot level ($\epsilon$). 
In such case, the density of states consists of two Andreev bound states 
branches separated by the pairing gap (Fig.\ \ref{figSAL}). For stronger
correlations, these subgap branches approach each other and can eventually 
cross at the singlet–doublet phase boundaries. 
The critical interaction for ground state transition at the half filling 
is $U=\Gamma_{S}$ (middle panel in Fig. \ref{figSAL}). 
 In the doublet regime ($U>\Gamma_{S}$), the subgap resonances cross each 
other, giving rise to a loop structure (bottom panel in Fig. \ref{figSAL}). 
In section \ref{sec:polarized}, we noticed that the states appearing in spin $\downarrow$ electron spectrum exhibit similar behavior (c.f.\ bottom panels in Fig. \ref{fig2}).  Presence of the zero-energy state in this spin channel did not affect significantly the characteristics of the singlet-doublet transition. In contrast, for spins directly coupled to nanochain, we note the qualitatively different  {\em avoided-crossing} behavior.

\bibliography{myBib}

\end{document}